# Power Up Consumption: Impact of Electric Vehicle Infrastructure on Household Finance

Sumit Agarwal[1], Yi Fan[2], Qiuxia Gao[3]

**Abstract**

To achieve carbon neutrality, electrifying vehicles in a cleaner electricity grid has become a key pathway for global cities. While extensive research has examined the environmental benefits of electric vehicles (EVs), their financial impacts remain underexplored. Using Singapore—a smart city-state—as a social laboratory, we combine high-resolution data on the geographic distribution of EV chargers with nearly 1 million individual-level monthly expenditure records from a major commercial bank and about 65,000 housing transaction records, spanning January 2019 to March 2021. Our results show that proximity to charging stations during the initial phase of EV infrastructure rollout significantly increases household consumption by 1–3%. This effect is driven primarily by public housing residents, who are generally more financially constrained than their private housing counterparts. We identify three channels underlying this increase in consumption: appreciation in housing wealth, catalysts for commercial activity and discretionary spending around EV infrastructure, and budget reallocation from fuel expenses to other consumption categories. Notably, hybrid car owners exhibit the strongest trade-offs—reducing fuel spending while increasing expenditure on entertainment or necessity goods among public housing residents. Overall, our findings demonstrate that green infrastructure can shape economic behavior and enhance consumer welfare, extending its impact beyond environmental objectives.

---

[1] Low Tuck Kwong Distinguished Professor of Finance, Economics and Real Estate, National University of Singapore; Email: bizagarw@nus.edu.sg

[2] Associate Professor, Department of Real Estate, NUS Business School, National University of Singapore; Email: yi.fan@nus.edu.sg

[3] PhD Candidate, Department of Real Estate, NUS Business School, National University of Singapore; Email: e0610864@u.nus.edu

Electrifying vehicles is a key pathway for decarbonizing urban transportation. To meet the global net-zero goal by 2050, cities intensive investment in electric vehicle (EV) infrastructure, particularly charging stations in public spaces and residential areas. Existing research focuses on environmental dimensions of EV transition, yet its socioeconomic ramifications——shaping housing choice, travel behavior, and household consumption patterns——remain underexplored. Understanding such socioeconomic dimensions is of particular importance, especially during the early stages of EV adoption, when public perceptions are under construction and behavioral responses remain highly adaptable. Using Singapore's staggered rollout of EV charging infrastructure in the early stages of EV adoption as a quasi-natural experiment, we investigate its impact on household finance including consumption, housing wealth, and local economic activities. Our findings shed light on the trajectory for other global cities embarking on their clean energy transitions.

A growing body of research in engineering and energy systems has examined the technical implications of EV infrastructure, particularly its role in optimizing battery production, managing electricity loads, improving grid operations, forecasting charging demand, and supporting vehicle routing efficiency (Powell et al. 2022; Cheng et al. 2024; Zhang et al. 2019; Aghajan-Eshkevari et al. 2023). In parallel, recent economics and policy studies have focused on disparities in access to EV charging—highlighting that rural areas, lower-income households, and racial minorities tend to face barriers to adoption due to limited infrastructure availability (Herberz et al. 2022; Taalbi and Nielsen 2021; Hanig et al. 2025; Lou et al. 2024; Steinbach and Blaschke 2024; Gherghina et al. 2025). Specifically, for lower-income households, the introduction of charging infrastructure may not only encourage EV adoption, but also influence everyday financial behaviors, as these households tend to be more sensitive to transportation and energy costs and thus more likely to exhibit behavioral shifts. Limited studies have explored the economic spillovers of EV infrastructure, such as Zheng et al. (2024) who find that public chargers can boost nearby commercial activity.

However, understanding how the EV infrastructure network can influence human behavior and underlying mechanisms is important. EV chargers, unlike many other forms of green infrastructure (e.g., solar panels or district heating), are not only climate assets, but also have socially visible and behaviorally interactive features of the urban landscape. Their presence can influence daily mobility, leisure choices, and local economic rhythms—making them a unique entry point for studying how green infrastructure can reshape everyday life. Such potential household-level behavioral changes remain largely absent from the current literature. Understanding them is crucial to uncovering the full social implications of the green transition.

Our study fills this gap by combining high-frequency individual-level consumption data from one of the largest commercial banks and detailed temporal and geographic information of EV charging infrastructure in Singapore (Figure S1). We begin by investigating the impacts of the EV charger installations on household consumption at the monthly level. To identify causal effects, we implement a difference-in-differences (DID) framework. Households in postal zones where at least one EV charger has been installed serve as the treatment group, while households in comparable areas without charger deployment form the control group. We find that the introduction of EV chargers leads to a significant increase in household consumption by 1.2% on average, controlling individual-, time- invariant, and time-varing unobservables. This effect is most pronounced at 1.9% among public housing residents, who are more budget-sensitive and more responsive to local infrastructure changes than their private housing counterparts. In addition, the impact of EV infrastructure on household consumption presents an increasing trend, reaching around 3% in the $20^{th}$ month after the charger installation.

To explain the findings, we show three evident mechanisms: (1) wealth effect, where being close to EV chargers raises perceived property values and expected capital gains, prompting households to spend more; (2) catalysts for commercial activity, as intensified charging network can stimulate business expansion and generate additional incidental neighborhood spending; and (3) petrol expenditure reallocation, where savings from reduced petrol costs free up household budgets for other discretionary purchases. A battery of heterogeneity analyses, based on housing type (public vs. private housing residents), transaction mode (debit/credit card, Paylah and Paynow, vs. Automated Teller Machine (ATM)/ on-site Network for Electronic Transfers Singapore (NETS)[4], spending category (necessities, luxury goods, entertainment, and petrol), and car ownership (individuals without cars, EV, hybrid, petrol car owners), support our channel establishment.

Our paper contributes to the studies on green infrastructure and household financial behaviors. First, we speak to the thin literature on the impact of EV charging infrastructure on household finance, by providing causal estimation. Burgeoning literature in engineering and energy systems focuses on the technical implications of EV infrastructure (Powell et al. 2022; Cheng et al. 2024; Zhang et al. 2019; Aghajan-Eshkevari et al. 2023). Recent economics and finance studies mainly investigate EV adoption and housing market outcomes (Greene et al. 2020; Javid et al. 2019; Liang et al. 2023; Globisch et al. 2019; Wang et al. 2019; Zheng et al. 2024). However, much less is known on how such EV infrastructure influences detailed household finance beyond housing and across time.

Our study also contributes to a broader literature on green infrastructure, such as rooftop solar installations which have been shown to reduce household electricity expenditures (Forrester et al. 2024; Deng and Newton 2017; Yozwiak et al. 2025). However, these studies largely focus on direct utility savings and do not capture broader behavioral responses, such as changes in household spending composition, budget reallocation, or discretionary consumption. Unlike solar panels, EV chargers are typically installed in public or semi-public spaces. They are highly visible, behaviorally interactive, and embedded in daily mobility patterns. As a result, EV charging infrastructure may influence not just energy-related spending, but also how households allocate time and money in their everyday routines. This makes it a particularly valuable, yet understudied, case for examining the behavioral spillovers of clean energy transitions.

---

[4] NETS has operated as a domestic payment network since 1985, while PayLah and PayNow are digital, real-time bank transfer systems introduced in 2014 and 2017, respectively. Details are provided in the Data section.

Last, we develop a mechanism-driven framework that links physical infrastructure to consumption responses through three distinct channels. This framing helps disentangle overlapping pathways and offers a template for evaluating other urban infrastructure expansion. We document heterogeneity in response across household types: the increase in consumption is more pronounced among public housing residents who are typically more budget-constrained, and hybrid car owners who are more flexible in shifting from brown to green energy sources. Such findings highlight the uneven distribution of infrastructure-induced benefits and underscore the importance of equity-conscious planning in sustainable urban development. Thus our results offer insights for designing clean energy investments that are not only environmentally effective but also socially inclusive.

# Results

## Impact of EV Infrastructure on Household Consumption

**Event Study and Parallel Pre-trend:** Literature on transport and energy shows that the spatial distribution of chargers is largely determined by engineering and regulatory constraints rather than household financial characteristics (Hall and Lutsey 2017; Andrenacci et al. 2016; Qian et al. 2020; Luo et al. 2015). According to Singapore Land Transport Authority (LTA) guidelines,[5] charger siting primarily depends on the availability of nearby grid capacity, proximity to existing substations and switchgear, and compliance with land-use and parking regulations. These criteria, set within national infrastructure planning frameworks, are driven by technical feasibility and safety standards. While road-network connectivity and land availability may correlate weakly with neighbourhood affluence, the underlying grid infrastructure that enables charger installation is largely predetermined and independent of household income distributions. In other words, the geographic distribution of public EV chargers is not likely endogenous to local income levels or socioeconomic patterns.

Based on the plausibly exogenous rollout of EV infrastructure network with respect to household

---

[5] Details of the report are available at: Land Transport Authority (LTA), *Guidelines for the Minimum EV Charging Provisions in Developments*

finance, we implement an event-study estimation following Eq. (1) in the Methods Section, to show temporal trend on the effect of EV charger installation on household consumption. Specifically, we manually collect installation date, latitude and longitude information of each EV charger installed between January 2019 and March 2021 during the early adoption phase in Singapore. Regarding household consumption, we obtain around 1 million individual monthly credit/debit card transaction and cash withdrawal records, from one of the largest commercial banks during the sample period (More details are provided in the Data Section). Figure 1 presents the trend analysis. While there is no material or statistically significant impact in the pre-treatment period, we observe a gradual and persistent increase in consumption after the EV charger installation. The estimates become statistically significant around month 12 and hover around 2–3%, equivalent to 37-55 USD. This pattern suggests a delayed but growing impact with the presence of EV chargers in the neighborhood on household spending.

**Average Local Effect:** The immaterial and statistically insignificant pre-trend estimates under the event study validate our DID specification. To estimate the average causal effect of EV infrastructure on household consumption, we implement the DID estimation following Eq. (2) in Methods Section. Table 1 presents the estimates. On average, there is 1.3% increase in household consumption after the EV charger installation in the neighborhood (Column (1)), including individual and year-month fixed effects but excluding inactive accounts with more than six consecutive months of zero transactions, following Agarwal and Qian (2014). To test the sensitivity of our findings, we progressively restrict the sample based on account activities. Columns (2) and (3) present estimates using refined definitions of inactive users—excluding individuals with zero transactions for more than three months or one month, respectively. The estimated effects remain stable and statistically significant, around 1.7%, suggesting that the observed increase in consumption reflects behavioral changes rather than data noise from inactive or extreme users. Specification in Column (4) further includes planning area-by-year fixed effects, controlling time-varying unobservables based on the same sample used in Column (3). This preferred estimate, taking into account the time-invariant and time-varying factors, remains at 1.2% (equivalent to 22 USD) and is statistically significant at a high 1% level. Considering around 1.7 million households by 2021 in Singapore, this number can be translated to 37.4 million USD across the city state.

**Robustness checks:** We further present a battery of robustness checks in Table 2, with baseline results shown in Column (1). In Column (2), we add back the top and bottom 1% of the trimmed consumption data. The estimate remains unchanged at 1.2% and is statistically significant at the 5% level. Results remain the same if using a more conservative clustering strategy at individual-by-year level (Column (3)). Column (4) presents findings by restricting the sample to chargers located in residential areas only, where household consumption is plausibly more responsive. By doing so, we alleviate concerns that stations in commercial or transit-oriented zones primarily serve non-local users, which may bias estimates of local consumption effects. Again, the estimated effect remains stable, both in magnitude and in level of statistical significance. In Column (5), we use an alternative dependent variable: the frequency of monthly consumption transactions, rather than total spending amount. We calculate the number of transactions per individual per month to capture behavioral shifts in spending intensity. The positive and statistically significant effect of 0.8% provides additional support that the observed behavioral shift is not limited to spending levels, but also reflects increased frequency of economic activity. Moreover, we move from incidence to intensity measure on EV infrastructure, by using the number of EV chargers in a postal zone (Column (6)). An additional EV charger leads to a 1% increase in consumption per household, equivalent to 22 USD. The estimate is statistically significant at the 5% level.

# Mechanism Analysis

We propose three channels to explain how EV charging infrastructure affects household consumption: (1) wealth effect, whereby rising property values near newly installed chargers increase households' perceived financial well-being and, in turn, their willingness to spend; (2) improved accessibility, as improved convenience and mobility encourage more frequent and incidental local purchases, particularly during waiting time to charge vehicles; and (3) reallocation from petrol spending, whereby expenditures previously devoted to petrol are redirected toward other forms of household consumption. Figure S2 illustrates the conceptual framework for mechanism analysis.

**Wealth effect:** We hypothesize a wealth effect to residents living close to the newly installed EV charging infrastructure and expect it is stronger for individuals with more stringent budget constraints. To test this hypothesis, we first conduct an event study following Eq. (1) to examine the impact of EV infrastructure rollout on housing prices among public and private housing residents, separately (Details of Singapore's two-layer housing market are presented in Supplementary Information A). Figure 2 shows that during the pre-treatment period, the differences in housing prices between the treatment group (in the vicinity of newly installed EV chargers) and control group (farther away from the EV chargers) are not statistically significant, supporting the parallel pre-trends assumption of DID estimation. Following the installation of EV chargers in the neighborhood, both public and private housing markets exhibit significant increases in local housing values. However, the impact in the public housing segment is delayed by approximately one year. In particular, this delayed but persistent increase reaches 5% in the $20^{th}$ month after charger installation, higher than the corresponding increase of 3% in the private housing segment. This lagged but larger response could be attributed to the tighter budget constraints and relatively lower car ownership rate among public housing residents, which make them less immediately responsive to the benefits of nearby EV infrastructure. In the longer term, the marginal benefit of improved access to chargers becomes greater for public housing residents, who face more limited mobility options and tend to be more sensitive to improvements in local amenities. The baseline estimates on the effect of charger installation on housing prices are presented in Table S3 with discussion in Supplementary Information C. We also conduct a battery of robustness checks as shown in Table S4, including alternative clustering levels, outlier removal, different control-group definitions, and the exclusion of executive condominiums. Across all specifications, the estimated effects remain stable in both magnitude and statistical significance.

Next, we compare the consumption patterns of public vs. private housing residents, using the same event study method as presented in Figure 3. While private housing residents do not show clear translation of appreciated housing wealth into consumption growth after EV charger installation, in contrast, public housing residents—who are likely more sensitive to perceived wealth gains—show a clear though delayed rise in consumption over time. This supports our hypothesis of housing wealth effect, which operates more strongly among the more budget-constrained population, such as the public housing residents.

We then extend the analysis to heterogeneity across public housing and find that differences in housing price responses that aligns closely with household wealth. Figure 4 shows that the introduction of nearby EV charging stations leads to large and statistically significant gains in the prices of 2-room and 3-room flats, generally occupied by lower-wealth households. Estimated effects are around 6% and 5%, respectively, at 1% significance level. In contrast, 4-room, larger, and executive units, which tend to house higher-income and higher-wealth families, exhibit near-zero and statistically insignificant changes. This gradient suggests that the welfare gains from EV-charging infrastructure are not evenly distributed but instead are concentrated among lower-wealth households. These households likely face greater frictions in vehicle ownership and mobility and thus place higher marginal value on improved charging accessibility.

**Catalysts for Commercial Activity:** The second hypothesized channel considers how EV chargers reshape local business environments. Beyond boosting commuters' spending while waiting to charge EVs, the charging infrastructure also likely stimulates the growth of nearby commercial establishments. By creating short windows for dining, shopping, or running errands while waiting, chargers generate additional foot traffic that can support new business formation. It may also improve amenity access for nearby non-drivers, reinforcing localized spending patterns. To examine this channel, we adopt a two-step strategy. First, we examine heterogeneity in household spending using bank transaction data. We focus on various payment methods and specifically ATM withdrawals at difference distance bands from chargers, as each payment channel in Singapore reflects distinct consumption patterns. Second, we apply Points of Interest (POI) data to analyze the evolving of commercial activities at postal-code zone levels, along with the staggered rollout of the charger installation.

We investigate household spending patterns with the DID framework as in the main analysis following Eq. (2). In Singapore, ATM cash withdrawals and NETS payments are typically associated with everyday neighborhood spending, such as food courts, convenience stores, and small daily essentials. Credit or debit cards tend to be used for higher-value purchases, while PayLah and PayNow are more common for online transfers.[6] Panel A of Table 3 shows that public housing residents exhibit significant increases in ATM and NETS spending after charger installation, with no corresponding rise in card or digital payments. This pattern suggests that EV chargers stimulate immediate, small-scale consumption among financially constrained households. Consistent with this interpretation, Panel B shows no significant spending changes among residents of private housing, who are generally less sensitive to localized retail opportunities.

We complement these results with a spatial heterogeneity analysis based on proximity to the nearest EV charger. Figure 5 depicts the estimated effect of charger installation on ATM withdrawals across 1-km distance bands for both public and private housing residents. A clear declining pattern emerges for public housing residents (Figure 5a): ATM withdrawals is as much as 5% higher within 1 km of a new charger, relative to those located more than 4 km away. The increase is smaller with 3% and 2% at 1–2 km and 2–3 km distance band, respectively. The effect dissipates beyond 3 km. This spatial gradient highlights the proximity-dependent nature of behavioral spillovers from EV infrastructure—greater charger accessibility appears to trigger more frequent neighborhood-level financial and consumption activity.

We then turn to the commercial landscape using the same DID estimation. Using OpenStreetMap, we identify all POIs in Singapore and compute the cumulative number of newly established POIs for each 4-digit postal code zone at the year-month level. Applying the same treatment and control definitions as in the bank-transaction analysis, we estimate how business formation responds to charger installation. Table 4 indicates that new EV chargers are associated with a marked increase in nearby commercial activities, although the effects vary by business types. At the aggregate level (Column 1), zones receiving chargers experience a 12.7% increase in newly established POI, suggesting that EV infrastructure can serve as a local economic catalyst. The event study shown in Figure 6 validates the parallel pre-trend assumption. Sectoral patterns reveal where this growth is concentrated. Dining and drinking establishments respond most strongly, with a 21.6% increase in new openings (Column 4), followed by retail and service businesses at 14.2% (Column 5). These sectors depend on short, convenience-driven demand, so the added foot traffic and dwell time from EV charging can help activate surrounding business and neighborhood development. In contrast, sectors such as entertainment, recreation, health, and medicine (Columns 2–3) show little response, likely due to higher entry costs and longer planning horizons. We also observe a decline in new public transit and facility-related POIs (Column 6), a category dominated by planned infrastructure such as bus stops and transit lines. Rather than displacement, this likely reflects zoning and land-use patterns, as chargers are more often installed in areas where transit infrastructure is already saturated.

---

[6] For more details on payment methods, refer to https://stripe.com/en-sg/resources/more/paynow-an-in-depth-guide and https://aspireapp.com/blog/different-payment-methods-for-your-e-commerce-business.

**Reallocation of petrol consumption:** With the rollout of EV infrastructure, vehicle consumption is also expected to change. At an extensive margin, new buyers are more likely to choose EVs or hybrid vehicles over petrol cars[7]. At an intensive margin, existing car owners tend to switch from petrol vehicles to electric or hybrid ones. If this hypothesis holds, the reduction in petrol-related spending would be reallocated to other types of consumption. We explicitly test it by differentiating households by vehicle ownership status (No car, EV owners, hybrid car owners, and petrol car owners) and housing type (public vs. private housing). Table 5 presents the estimates.

Among public housing residents, who are typically more budget constrained, we observe clear evidence of consumption reallocation following EV charger installation. Among vehicle owners, hybrid car users experience a statistically significant decline in fuel spending of approximately 8.7%, alongside increases of 4.5% in necessity spending and 5.1% in entertainment spending. These results suggest that fuel cost savings are partially reallocated toward both essential goods and discretionary consumption, reflecting an adjustment in household spending portfolios following reduced reliance on petrol. In contrast, individuals that already owned electric vehicles show no significant change in fuel expenditures and no statistically significant responses in either necessity or entertainment spending.

Interestingly, households without car ownership also exhibit consumption responses following charger installation, with modest but statistically significant increases in both necessity and entertainment spending. This pattern is consistent with the second channel, whereby the expansion of nearby commercial activities induced by EV charging infrastructure stimulates local consumption even among residents without vehicles nearby. Overall, the installation of EV charging infrastructure reallocates household spending by reducing fuel expenditures, particularly for hybrid vehicle owners, and indirect spillovers to non-car households through increased neighborhood commercial activity.

Among private housing residents, who generally have greater financial flexibility, consumption responses follow a distinct pattern. Hybrid vehicle owners exhibit a substantial reduction in fuel spending of approximately 26.4%—a markedly larger response than that observed in public housing—while reallocating only a modest share of these savings toward necessity or entertainment consumption. EV owners display the strongest behavioral response, with pronounced increases in discretionary spending, including a 35.2% rise in luxury consumption and a 20.1% increase in entertainment spending. Conversely, individuals without car ownership experience significant reductions in luxury and entertainment spending, suggesting a potential shift toward saving rather than immediate consumption. Taken together, these patterns indicate that wealthier households are more likely to translate EV adoption into lifestyle upgrading rather than treating energy savings purely as a cost-saving adjustment.

---

[7] EVs in Singapore generally have similar initial purchase costs about S$130,000, but lower running and maintenance costs relative to petrol cars, according to AION

# Discussion

We present pioneering evidence on how EV charging infrastructure shapes household finance, drawing on Singapore's quasi-natural experiment of staggered rollout during its early-stage EV transition. Using a DID framework and rich micro-level data including geolocated charger deployments, housing transactions, and individual financial records, we find that the installation of EV chargers leads to significant increase in household spending, particularly among public housing residents, who are more financially constrained. These effects are consistent with three interrelated mechanisms: perceived wealth gains from rising property values, expanded business activity that supports greater neighborhood consumption, and savings from reduced petrol expenditure.

Our findings reveal that EV charging infrastructure delivers not only environmental gains but also tangible socioeconomic spillovers, particularly for lower-income communities. These distributional effects carry important policy implications. Current planning and engineering frameworks for charger deployment largely emphasize technical optimization, such as network coverage, load balancing, and vehicle range, while overlooking user-side behavior. Our results show that incorporating household budget sensitivity and neighborhood economic context into siting decisions can substantially enhance social welfare. Moreover, the stronger consumption responses observed in public-housing neighborhoods suggest that chargers can act as localized economic catalysts, supporting small businesses, enhancing perceived asset value, and improving quality of life. To fully realize these potential benefits, infrastructure planning should adopt a more integrated approach with community inclusive development objectives. Policy tools such as targeted subsidies and zoning flexibility can guide charger deployment to maximize the social-equity impact of public investment.

Beyond Singapore, these insights are highly relevant for economies where EV is still emerging. The early stage of EV infrastructure rollout in Singapore serves as a valuable social laboratory for other economies especially developing ones, where the EV transition remains in its infancy: adoption rates are low, charging networks sparse, and infrastructure expansion still heavily state-driven. In addition, the residential structures in these contexts often resemble Singapore's public housing model: large-scale, government-built estates accommodating much of the lower- and middle-income population. To assess external relevance, we compile EV-infrastructure indicators for Malaysia, Indonesia, Thailand, Brazil, Colombia, India, and Mexico as of 2024 and benchmark them against Singapore's 2021 experience in Table 6. In 2021, Singapore had roughly 2,000 public charging points, 3.7 chargers per 10,000 people with around 20% classified as fast chargers and an EV-adoption rate of 3.8%. These levels were below the global average across all three dimensions and mirror 2024 conditions in several developing economies: Malaysia (1.43 chargers per 10,000 people; 4% adoption), Indonesia (0.11; 7%), and Brazil (0.60; 6.5%). Singapore's early configuration—moderate charger density, limited fast-charging share, thus provides a useful analogue for the formative stages of electrification now underway across much of the Global South.

Singapore's experience illustrates how the introduction of public EV charging infrastructure can have influence beyond environmental consequences, but also ramifications into household finance and neighborhood economic activity. The observed increase in local spending near chargers, particularly among lower- and middle-income households, suggests that public charging points can serve as catalysts at the community level, stimulating local commerce and social interaction while allowing cleaner mobility. These dynamics are likely to extend beyond Singapore, as many developing economies with dense residential clusters and limited access to home-based charging may experience similar community-level benefits once shared public chargers become integrated into daily life. Importantly, these patterns also expose a distributional effects of the energy transition. Proximity to public chargers disproportionately benefits lower- and middle-income households, positioning public infrastructure as a potential economic equalizer. As developing economies are expanding their EV networks, ensuring equitable spatial access will be vital to prevent the clean-mobility transition from reinforcing existing inequalities. Singapore's case shows that when charging investments are inclusive in design and equitable in siting, the benefits of decarbonized transport can support local economies, improving access to sustainable mobility, and advancing a more just and inclusive development.

Our analysis has several limitations that point to promising directions for future research. First, although our banking data provide rich visibility into household spending, they do not include charger-level EV-charging transactions. Access to billing records or telematics-based energy-use logs would allow future work to more cleanly isolate charging-related expenditures and refine the mechanisms we document. Second, we lack high-resolution mobility data, preventing us from observing whether improved charging access alters travel frequency, routing, or time spent around charging sites. Incorporating anonymized mobile-phone or GPS traces would shed light on these behavioral adjustments. Third, our data do not capture vehicle adoption, residential moves, or longer-run sorting responses, which may unfold as the charging network expands. Linking administrative vehicle registrations, housing mobility data, or panel surveys would help quantify these dynamic channels and provide a fuller picture of the welfare consequences of electrification infrastructure.

# Method

## Data

**EV infrastructure** Singapore's national EV infrastructure plan was launched in 2020, with a goal of deploying 60,000 charging points—40,000 in public car parks and 20,000 in private premises-by 2030. The Land Transport Authority (LTA) has prioritized deployment in public housing estates (HDB car parks), with tenders awarded to ensure wide geographic coverage and accessibility (Detailed background in Supplementary Information A). We manually compile installation date, latitude and longitude data of 129 EV chargers installed between January 2019 and March 2021. As shown in Figure S1, a majority of early-stage chargers began operating between 2019 and 2021 and are distributed across the island with higher densities in the central and eastern regions. Our manually constructed EV infrastructure dataset also includes land use types and charger ownership. Chargers are installed near diverse land use types including public housing (about 40%), private residences like condominiums (20%), commercial facilities (e.g., malls and offices), schools, and industrial parks—reflecting broad integration of EV infrastructure across residential, commercial, and institutional settings, with the highest concentration in residential areas.[8] Three main owners—Bluecharge, Shell Recharge, and SP Mobility—operate the chargers. Bluecharge, the dominant provider, concentrates its installations in the eastern and central regions, whereas Shell Recharge and SP Mobility adopt a more dispersed pattern, with stations located more in residential and mixed-use areas.

**Household finance** We obtain monthly consumption records of 230,000 individuals from a proprietary dataset provided by one of the largest commercial banks in Singapore between January 2019 and March 2021. Singapore's financial system is highly digitized and consolidated, with a small number of domestic banks providing comprehensive coverage of the resident population. As one of the largest banks in Singapore, it operates more than twice as many branches and over four times as many ATMs as other major banks, serving more than 80% of resident households (Agarwal and Qian 2014). Therefore, the transaction data we use is highly representative of household spending patterns in Singapore. The dataset includes transaction date, amount, merchant category, and payment type (credit cards, debit cards, NETS accounts, PayLah, PayNow, and ATM). These payment types are informative because they map onto distinct types of consumption in Singapore. ATM cards and the NETS system remain widely used for everyday, in-person transactions such as payments at food courts, hawker centers, and small retail outlets. Credit and debit cards are typically preferred for larger retail purchases, dining, and online shopping. Differently, PayLah and PayNow function as fintech platforms linked to users' bank accounts, and are primarily employed for digital transfers, such as peer-to-peer transfer, casual exchanges like splitting bills or online transfer to service providers.

[8] EV chargers in our data are dominated by publicly accessible infrastructures. In Singapore, only residents of landed properties (approximately 5%) are allowed to install private-home chargers through their car dealers. The number of condominiums equipped with private charging points remains very small during our study period. A majority of households rely on public EV charging stations. Refer to https://www.propertyguru.com.sg/property-guides/electric-car-singapore-condo-ev-charging-48739 for more details.

We clean the data by first removing outliers, defined as the top and bottom 1% of expenditures within each payment category (cards, ATM, NETS, PayNow, and PayLah). We then aggregate the cleaned transactions across categories to obtain individual-level monthly total expenditure, and excluding the nonactive accounts following the literature (Agarwal and Qian 2014). Besides, we retain the data for individuals aged 22 to 65, as this age range generally represents the economically active working population. Table S1 reports summary statistics for the variables used in the consumption analyses. Private housing residents spend more on average each month (SGD 2,777; approximately USD 2,154), compared to the public housing counterparts (SGD 2,312; USD 1,793). A larger proportion of private housing residents are credit/debit card users, whereas public housing residents rely more on ATM/NETS payments. Demographics across both groups are generally comparable.

**Housing transaction** Singapore's housing market is characterized by a distinct two-tier structure. Over 80% of residents live in public housing provided by the government housing authority Housing & Development Board (HDB), which targets lower- and middle-income households through eligibility rules such as income ceilings and minimum occupancy periods. In contrast, private housing—including condominiums and landed properties—caters to higher-income groups and faces fewer regulatory restrictions (Agarwal et al.2025) (Detailed background in Supplementary Information B). We collect transaction records of public and private housing in the sample period from the HDB and the Urban Redevelopment Authority (URA). The housing data includes transaction date, postal code, transaction price, and housing characteristics such as floor area, number of units, type of housing and age.

We geocoded using the OneMap API to obtain latitude and longitude of each building, to match with EV chargers. To define treatment and control groups, we classify buildings within 1 km of the earliest-start-date chargers as the treatment group follow the method of Liang et al. (2023)and those in 1-6 km without overlapping with treatment groups as the control group. Table S2 reports summary statistics for the variables used in the housing analyses. The average transaction price is approximately SGD 1.4 million (USD 1.08 million) for private housing and SGD 442,000 (USD 342,796) for public housing. Private housing units are newer, with an average of 9 years, while public flats average 31 years.

**Points of Interest (POI) Data**: We obtained Singapore's POI data by scraping business and facility records from OpenStreetMap (OSM). The raw POI records contain each establishment's latitude and longitude, opening date, and detailed OSM category (art entertainment, sport recreation, dining and drinking, retail, business service, health and medicine, public transit facility, etc.). We first aggregate these fine-grained categories into broader commercial types: Entertainment & Recreation, Health & Medicine, Dining & Drinking, Retail & Service, and Public Transit & Facility. To align POIs with the analytical units used in the bank-transaction analysis, we employ the OneMap API to map each POI's coordinates to the corresponding 4-digit postal code zone and merge these records to the existing panel of treated and control zones. For every postal zone, we then compute the cumulative number of newly established POIs at the year-month level from January 2019 to March 2021, generating a granular measure of monthly business formation.

## Empirical Strategy on Consumption

**Event-study:** We apply an event-study method to evaluate the monthly impact of EV charger installation on household consumption:

$$Ln(Y_{imt}) = \beta_0 + \sum_{k=-2}^{-10} \theta^k \left(Treat_{im} \times Pre_{mt}^k\right) + \sum_{k=0}^{20} \omega^k \left(Treat_{im} \times Post_{mt}^k\right) + \eta Ln(Income_{it}) + \delta_i + \tau_t + \gamma_{rt} + \epsilon_{imt} \quad (1)$$

Where $Y_{imt}$ is the consumption amount of individual $i$ located in zone $m$ (4-digit postal code) in year-month $t$. $Treat_{im}$ is a binary variable equal to 1 if an individual $i$ resides in treated zone $m$ with EV chargers, and 0 for individuals in control group. Specifically, the treatment group is defined as zones with EV chargers at the 4-digit postal code level.[9] The control group consists of other 4-digit postal code zones within the same 3-digit postal code region of the treatment group but without EV chargers. If there is more than one EV charger in a 4-digit postal code zone, we use the earliest start date of chargers as the treatment time. $Pre_{mt}^k$ ($Post_{mt}^k$) is a binary variable that takes the value of 1 in the $k$-month before (after) the EV charger installation. We include monthly income $Income_{it}$ as a control variable, as it is a major determinant for consumption. $\delta_i$ captures individual fixed effects, accounting for time-invariant individual unobservable factors. $\tau_t$ denotes year-month fixed effects, which flexibly control for time trends and common temporal shocks that affect all individuals. Moreover, we include $\gamma_{rt}$, a set of planning area-by-year fixed effects, to control for neighborhood-level differences, and other regional time-varying unobserved confounders that may influence household spending. Standard errors are clustered at individual and month levels.

[9] In Singapore, a six-digit postal code represents one building. The first three digits of a six-digit postal code correspond to the postal region, which generally aligns with a town planning center that typically comprises between 4,000 and 7,000 residential units. There are approximately 400 such town centers across the country. At a finer granularity, the first four digits define the postal code zone, each containing around 500 to 1,000 residential units, with a total of roughly 2,400 such zones nationwide.

**Baseline Estimation:** The DID estimation that we adopt to estimate the average causal effect of EV charger installation on household consumption in the neighborhood is specified as follows:

$$Ln(Y_{imt}) = \beta_0 + \beta_1 Treat_{im} * Post_{mt} + \eta Ln(Income_{it}) + \delta_i + \tau_t + \gamma_{rt} + \epsilon_{imt} \quad (2)$$

where $Post_{mt}$ equals 1 for year-month $t$ after the earliest start date of the EV charger in zone $m$. Definitions for other variables remain the same as those for Eq.(1).

## Empirical Strategy on Housing Prices

**Event-study:** The event study used to estimate the impact of EV charger installation on monthly housing price is specified as below:

$$Ln(Y_{ijt}) = \beta_0 + \sum_{k=-2}^{-10} \theta^k \left(Treat_{ij} \times Pre_{jt}^k\right) + \sum_{k=0}^{20} \omega^k \left(Treat_{ij} \times Post_{jt}^k\right) + \delta X_{it} + \theta_{ij} + \tau_t + \gamma_{rt} + \epsilon_{ijt} \quad (3)$$

Where $Y_{ijt}$ is the transaction price of house in building (postal code) $i$ near charger $j$ (within 6 km) in year-month $t$. $Treat_{ij}$ is a binary variable that equals 1 if the housing unit $i$ locates within 1 km of EV charger $j$, following the specification in Liang et al. (2023). It is equal to 0 if locating 1-6 km away from charger $j$ but not in the treatment group of any other charger. To address potential bias arising from overlapping treatment issue that some housing units may locate within 1 km of multiple EV chargers installed at different time, we retain housing observations in the treatment group associated with the first-built charger. We adopt 6 km as the control-group boundary, corresponding to the maximum distance observed between any building and the nearest charger in our data. As a robustness check, we also form control groups based on an alternative distance threshold and report the result in the robustness table. $Pre_{jt}^k$ and $Post_{jt}^k$ are binary variables that take the value of 1 in the $k$-month before and after the EV charger installation, respectively.

We include a set of control variables on housing attributes, $X_{it}$, that affect housing prices, such as floor area, number of units, and age of buildings. Additionally, we incorporate a rich set of fixed effects to account for unobserved spatial and temporal confounders. Specifically, $\tau_t$ denotes time fixed effects (year and month), controlling for variations common to all units at a given time, such as macroeconomic conditions or seasonal consumption trends. Considering some buildings may fall within the control group of multiple chargers, we include postal code-by-charger fixed effects, $\theta_{ij}$. These fixed effects also absorb any time-invariant differences across localized areas associated with a specific EV charger, such as amenity or neighborhood characteristics. In addition, $\gamma_{rt}$ represents planning area-by-year fixed effects, which account for time-varying shocks such as local changes or infrastructure investments that could differentially affect property values across regions and over time. Standard errors are clustered at the postal code level.

**Baseline Estimation:** Related, the DID estimation used to examine the average effect of the staggered rollout of EV infrastructure on housing price is specified as below:

$$Ln(Y_{ijt}) = \beta_0 + \beta_1 Treat_{ij} \times Post_{jt} + \delta X_{it} + \theta_{ij} + \tau_t + \gamma_{rt} + \epsilon_{ijt} \quad (4)$$

where $Post_{jt}$ equals one if the transaction date $t$ occurs after the installation of EV charger $j$. Other variables follow the same definitions as in Eq. (3).

## References

Agarwal, Sumit, and Wenlan Qian. 2014. "Consumption and Debt Response to Unanticipated Income Shocks: Evidence from a Natural Experiment in Singapore." *American Economic Review* 104 (12): 4205–30.

Aghajan-Eshkevari, Saleh, Mohammad Taghi Ameli, and Sasan Azad. 2023. "Optimal Routing and Power Management of Electric Vehicles in Coupled Power Distribution and Transportation Systems." *Applied Energy* 341: 121126.

Andrenacci, Natascia, Roberto Ragona, and Gaetano Valenti. 2016. "A Demand-Side Approach to the Optimal Deployment of Electric Vehicle Charging Stations in Metropolitan Areas." *Applied Energy* 182: 39–46.

Cheng, Anthony L, Erica RH Fuchs, and Jeremy J Michalek. 2024. "US Industrial Policy May Reduce Electric Vehicle Battery Supply Chain Vulnerabilities and Influence Technology Choice." *Nature Energy* 9 (12): 1561–70.

Deng, Gary, and Peter Newton. 2017. "Assessing the Impact of Solar PV on Domestic Electricity Consumption: Exploring the Prospect of Rebound Effects." *Energy Policy* 110: 313–24.

Forrester, Sydney P, Cristina Crespo Montañés, Eric O'Shaughnessy, and Galen Barbose. 2024. "Modeling the Potential Effects of Rooftop Solar on Household Energy Burden in the United States." *Nature Communications* 15 (1): 4676.

Gherghina, Mircea, Fedor A Dokshin, and Benjamin Leffel. 2025. "Unequal Solar Photovoltaic Performance by Race and Income Partly Reflects Financing Models and Installer Choices." *Nature Energy*, 1–10.

Globisch, Joachim, Patrick Plötz, Elisabeth Dütschke, and Martin Wietschel. 2019. "Consumer Preferences for Public Charging Infrastructure for Electric Vehicles." *Transport Policy* 81: 54–63.

Greene, David L, Eleftheria Kontou, Brennan Borlaug, Aaron Brooker, and Matteo Muratori. 2020. "Public Charging Infrastructure for Plug-in Electric Vehicles: What Is It Worth?" *Transportation Research Part D: Transport and Environment* 78: 102182.

Hall, Dale, and Nic Lutsey. 2017. "Emerging Best Practices for Electric Vehicle Charging Infrastructure." *The International Council on Clean Transportation (ICCT): Washington, DC, USA* 54.

Hanig, Lily, Catherine Ledna, Destenie Nock, et al. 2025. "Finding Gaps in the National Electric Vehicle Charging Station Coverage of the United States." *Nature Communications* 16 (1): 561.

Herberz, Mario, Ulf JJ Hahnel, and Tobias Brosch. 2022. "Counteracting Electric Vehicle Range Concern with a Scalable Behavioural Intervention." *Nature Energy* 7 (6): 503–10.

Javid, Roxana J, Mahmoud Salari, and Ramina Jahanbakhsh Javid. 2019. "Environmental and Economic Impacts of Expanding Electric Vehicle Public Charging Infrastructure in California′ s Counties." *Transportation Research Part D: Transport and Environment* 77: 320–34.

Liang, Jing, Yueming Qiu, Pengfei Liu, Pan He, and Denise L Mauzerall. 2023. "Effects of Expanding Electric Vehicle Charging Stations in California on the Housing Market." *Nature Sustainability* 6 (5): 549–58.

Lou, Jiehong, Xingchi Shen, Deb A Niemeier, and Nathan Hultman. 2024. "Income and Racial Disparity in Household Publicly Available Electric Vehicle Infrastructure Accessibility." *Nature Communications* 15 (1): 5106.

Luo, Chao, Yih-Fang Huang, and Vijay Gupta. 2015. "Placement of EV Charging Stations—Balancing Benefits Among Multiple Entities." *IEEE Transactions on Smart Grid* 8 (2): 759–68.

Powell, Siobhan, Gustavo Vianna Cezar, Liang Min, Inês ML Azevedo, and Ram Rajagopal. 2022. "Charging Infrastructure Access and Operation to Reduce the Grid Impacts of Deep Electric Vehicle Adoption." *Nature Energy* 7 (10): 932–45.

Qian, Tao, Chengcheng Shao, Xuliang Li, Xiuli Wang, and Mohammad Shahidehpour. 2020. "Enhanced Coordinated Operations of Electric Power and Transportation Networks via EV Charging Services." *IEEE Transactions on Smart Grid* 11 (4): 3019–30.

Steinbach, Sarah A, and Maximilian J Blaschke. 2024. "How Grid Reinforcement Costs Differ by the Income of Electric Vehicle Users." *Nature Communications* 15 (1): 9674.

Taalbi, Josef, and Hana Nielsen. 2021. "The Role of Energy Infrastructure in Shaping Early Adoption of Electric and Gasoline Cars." *Nature Energy* 6 (10): 970–76.

Wang, Bo, Payman Dehghanian, Shiyuan Wang, and Massimo Mitolo. 2019. "Electrical Safety Considerations in Large-Scale Electric Vehicle Charging Stations." *IEEE Transactions on Industry Applications* 55 (6): 6603–12.

Yozwiak, Madeline, Galen Barbose, Sanya Carley, et al. 2025. "The Effect of Residential Solar on Energy Insecurity Among Low-to Moderate-Income Households." *Nature Energy*, 1–12.

Zhang, Yongmin, Jiayi Chen, Lin Cai, and Jianping Pan. 2019. "Expanding EV Charging Networks Considering Transportation Pattern and Power Supply Limit." *IEEE Transactions on Smart Grid* 10 (6): 6332–42.

Zheng, Yunhan, David R Keith, Shenhao Wang, Mi Diao, and Jinhua Zhao. 2024. “Effects of Electric Vehicle Charging Stations on the Economic Vitality of Local Businesses.” *Nature Communications* 15 (1): 7437.

## Tables and Figures

**Table 1:** Baseline Estimate on Total Consumption

| | (1) | (2) | (3) | (4) |
|---|---|---|---|---|
| $\boldsymbol{Treat * Post}$ | $0.013^{***}$ | $0.016^{***}$ | $0.017^{***}$ | $0.012^{**}$ |
| | (0.005) | (0.005) | (0.005) | (0.005) |
| Observations | 1241564 | 1191441 | 1089378 | 1089378 |
| R2 | 0.54 | 0.53 | 0.53 | 0.53 |
| Mean Dep Variable | 2103.97 | 2144.71 | 2240.86 | 2240.86 |
| Individual FE | Y | Y | Y | Y |
| Year-month FE | Y | Y | Y | Y |
| Planning area-by-Year FE | | | | Y |
| Nonactive transactions | $\leq 6 months$ | $\leq 3 months$ | $\leq 1 month$ | $\leq 1 month$ |

Notes: Standard errors are clustered at the individual and month level, reported in parentheses. *$P < 0.10$; **$P < 0.05$; ***$P < 0.01$; Dependent variables are the logarithmic values of monthly consumption amounts from January 2019 to March 2021. Total consumption is constructed by aggregating across payment types after trimming the top and bottom 1% of expenditures within each type. The logarithmic value of monthly income is included as a control variable. $Treat$ is a binary variable equal to one for individuals residing in four-digit postal code zones with EV chargers (treatment group), and zero for those in other four-digit zones within the same three-digit region without chargers (control group). $Post$ is a binary variable equal to one for consumption recorded after the installation of the EV charger. Columns (1)–(3) exclude nonactive accounts based on the maximum number of consecutive months with zero consumption, using thresholds of more than six months, three months, and one month, respectively.

**Table 2:** Robustness Check on Total Consumption

| (1) | (2) | (3) | (4) | (5) | (6) |
|---|---|---|---|---|---|
| Baseline | Full Sample | Different Cluster | Subsample | Alternative Dep. Var | Treatment Intensity |
| Figure/robustness_dbs | | | | | |

Notes: Standard errors are clustered at the individual and month level, reported in parentheses. *P < 0.10; **P < 0.05; ***P < 0.01; Dependent variables are the logarithmic values of monthly consumption amounts from January 2019 to March 2021, otherwise stated in the column. Consumption amount is constructed by aggregating across payment types after trimming the top and bottom 1% of expenditures within each type. The logarithmic value of monthly income is included as a control variable. $Treat$ is a binary variable equal to one for individuals residing in four-digit postal code zones with EV chargers (treatment group), and zero for those in other four-digit zones within the same three-digit region without chargers (control group). $Post$ is a binary variable equal to one for consumption recorded after the installation of the EV charger. We exclude inactive accounts with more than one month of zero consumption. Individual, year-month, and planning area-by-year fixed effects are included. Column (2) includes the top and bottom 1% outliers to keep the full sample data. Column (3) clusters standard errors at the individual-by-year level. Column (4) restricts the sample of charging stations located in residential areas. Column (5) uses an alternative dependent variable of monthly consumption frequency, defined as the total number of spending transactions per individual per month. Column (6) replaces the binary treatment with a continuous measure based on the number of EV chargers in each region.

**Table 3:** Heterogeneity in Consumption by Housing Type and Payment Method

| | (1) All | (2) Credit&Debit Card | (3) ATM&Nets | (4) Paylah& Paynow |
|---|---|---|---|---|
| **Panel A: Public Housing** | | | | |
| $\boldsymbol{Treat * Post}$ | 0.012** | -0.002 | 0.013** | -0.000 |
| | (0.005) | (0.007) | (0.006) | (0.012) |
| Observations | 898851 | 625705 | 793672 | 349596 |
| R2 | 0.53 | 0.63 | 0.49 | 0.44 |
| Mean Dep Variable | 2208.04 | 926.97 | 1234.08 | 983.86 |
| **Panel B: Private Housing** | | | | |
| $\boldsymbol{Treat * Post}$ | -0.016 | -0.033 | 0.010 | -0.055 |
| | (0.021) | (0.024) | (0.028) | (0.047) |
| Observations | 81545 | 76193 | 60274 | 39453 |
| R2 | 0.56 | 0.64 | 0.48 | 0.40 |
| Mean Dep Variable | 3093.77 | 1471.45 | 1030.33 | 952.91 |

Notes: Standard errors are clustered at the individual and month level, reported in parentheses. *P $<$ 0.10; **P $<$ 0.05; ***P $<$ 0.01; Dependent variables are the logarithmic values of monthly consumption amounts from January 2019 to March 2021. Consumption amount is constructed by aggregating across payment types after trimming the top and bottom 1% of expenditures within each type. We include individual, year-month, planning area-by-year fixed effects and the logarithmic value of monthly income as the control variable. $Treat$ is a binary variable equal to one for individuals residing in four-digit postal code zones with EV chargers (treatment group), and zero for those in other four-digit zones within the same three-digit region without chargers (control group). $Post$ is a binary variable equal to one for consumption recorded after the installation of the EV charger. We exclude inactive accounts with more than one month of zero consumption.

**Table 4:** Heterogeneous Impacts of Nearby POIs

| | (1) | (2) | (3) | (4) | (5) | (6) |
|---|---|---|---|---|---|---|
| | Overall | Entertainment & Recreation | Health& Medicine | Dining & Drinking | Retail & Service | Public Transit & Facility |
| $\boldsymbol{Treat * Post}$ | 0.119*** | 0.131 | 0.053 | 0.196*** | 0.133** | -0.162** |
| | (0.040) | (0.087) | (0.100) | (0.052) | (0.058) | (0.077) |
| Percentage | 12.67 | 14.01 | 5.48 | 21.62 | 14.20 | -14.97 |
| Observations | 8226 | 3006 | 1326 | 5826 | 5649 | 4083 |
| R2 | 0.79 | 0.44 | 0.47 | 0.70 | 0.69 | 0.32 |
| Mean Dep Variable | 0.38 | 0.09 | 0.09 | 0.26 | 0.17 | 0.07 |

Notes: Standard errors are clustered at the 4-digit postal code and month level, reported in parentheses. *P $<$ 0.10; **P $<$ 0.05; ***P $<$ 0.01; Dependent variables are the number of newly established POIs at the 4-digit postal-code × year-month level from January 2019 to March 2021. Because the dependent variable contains many zeros, all regressions are estimated using PPML. $Percentage$ reports the percentage change in the dependent variable implied by the PPML coefficient. $Treat$ is a binary variable equal to one for four-digit postal code zones with EV chargers (treatment group), and zero for other four-digit zones within the same three-digit region without chargers (control group). $Post$ is a binary variable equal to one for time after the installation of the EV charger. In Column (6), public transit and facilities refer to bus stops, bus lines, MRT transit hubs, and other community amenities.

**Table 5:** Heterogeneity in Consumption by Housing Type, Consumption Category and Car Ownership

| | (1) | (2) | (3) | (4) |
|---|---|---|---|---|
| | Fuel Spending | Luxury Spending | Necessity Spending | Entertainment Spending |
| | **Panel A: Public Housing** | | | |
| $\boldsymbol{NoCar * Post}$ | 0.000 | -0.013 | 0.045* | 0.057* |
| | (.) | (0.038) | (0.025) | (0.033) |
| $\boldsymbol{EVOwner * Post}$ | 0.000 | 0.015 | -0.026 | -0.044 |
| | (.) | (0.037) | (0.024) | (0.032) |
| $\boldsymbol{HybridEV * Post}$ | -0.087*** | 0.009 | 0.045*** | 0.051*** |
| | (0.030) | (0.021) | (0.013) | (0.017) |
| Observations | 96556 | 236446 | 401800 | 359415 |
| R2 | 0.63 | 0.55 | 0.54 | 0.45 |
| Mean Dep Variable | 205.40 | 566.33 | 364.66 | 507.05 |
| | **Panel B: Private Housing** | | | |
| $\boldsymbol{NoCar * Post}$ | 0.000 | -0.382*** | -0.091 | -0.336*** |
| | (.) | (0.125) | (0.084) | (0.102) |
| $\boldsymbol{EVOwner * Post}$ | 0.000 | 0.352*** | 0.067 | 0.201** |
| | (.) | (0.117) | (0.076) | (0.092) |
| $\boldsymbol{HybridEV * Post}$ | -0.264*** | 0.071 | -0.007 | 0.036 |
| | (0.077) | (0.067) | (0.049) | (0.058) |
| Observations | 21684 | 41885 | 67726 | 60940 |
| R2 | 0.62 | 0.50 | 0.60 | 0.51 |
| Mean Dep Variable | 211.86 | 805.57 | 552.74 | 836.92 |

Notes: Standard errors are clustered at the individual and month level, reported in parentheses. *$P < 0.10$; **$P < 0.05$; ***$P < 0.01$; Dependent variables are the logarithmic values of monthly consumption amounts after removing the top and bottom 1% outliers, from January 2019 to March 2021. We include all fixed effects and the logarithmic value of monthly income as the control variable. $NoCar$, $EVOwners$ and $HybridEV$ represent treatment groups with different fuel consumption behaviors. Specifically, $NoCar$ indicates individuals with no car ownership, defined by the absence of both fuel purchases and automotive-related expenditures (e.g., insurance, maintenance, or parking fees) during the sample period. $EVOwners$ refers to individuals who owned electric vehicles (exhibited no recorded fuel purchases throughout the sample period, but displayed evidence of vehicle ownership through automotive-related spending) or individuals who switched to EVs (previously had fuel purchases but exhibited no recorded fuel spending for at least six consecutive months by March 2021). $HybridEV$ refers to individuals who owned hybrid electric vehicles but did not fully switch to EVs. These individuals did not experience a sustained period of six or more consecutive months without fuel spending by March 2021 and displayed intermittent fuel purchases throughout the 27-month sample period, with fuel consumption recorded in fewer than 12 months. We exclude inactive accounts with more than one month of zero consumption.

**Table 6:** Summary on the Electric vehicle Charging Infrastructure and Adoption

| Country<br><br>Year | Total Public<br><br>Charging Points | Chargers per<br><br>10,000 People | DC Fast-Charger<br><br>Share (%) | EV Adoption<br><br>Rate (%) |
|---|---|---|---|---|
| Singapore 2021 | 2,000 | 3.670 | 20.00 | 3.8 |
| Thailand 2024 | 11,467 | 1.601 | 50.42 | 13 |
| Malaysia 2024 | 1430 | 0.397 | 19.16 | 4 |
| Indonesia 2024 | 3,202 | 0.112 | 27.00 | 7 |
| Brazil 2024 | 12,700 | 0.597 | - | 6.5 |
| Colombia 2024 | 389 | 0.073 | - | 7.5 |
| India 2024 | 40,000 | 0.273 | 35.00 | 2 |
| Mexico 2024 | 3,212 | 0.243 | 16.75 | 8.2 |
| Global average 2024 | 27,179 | 6.500 | 22.00 | 25 |

Notes: This table compares the scale and measures of EV charging infrastructure across Singapore (2021) and major developing economies as of 2024. Data on the total public charging points and the DC fast-chargers share are compiled from media reports and the Roland Berger Global EV Outlook. Population figures are obtained from the World Population Review, and we compute the chargers per 10,000 people. EV penetration rates are sourced from the International Energy Agency report.

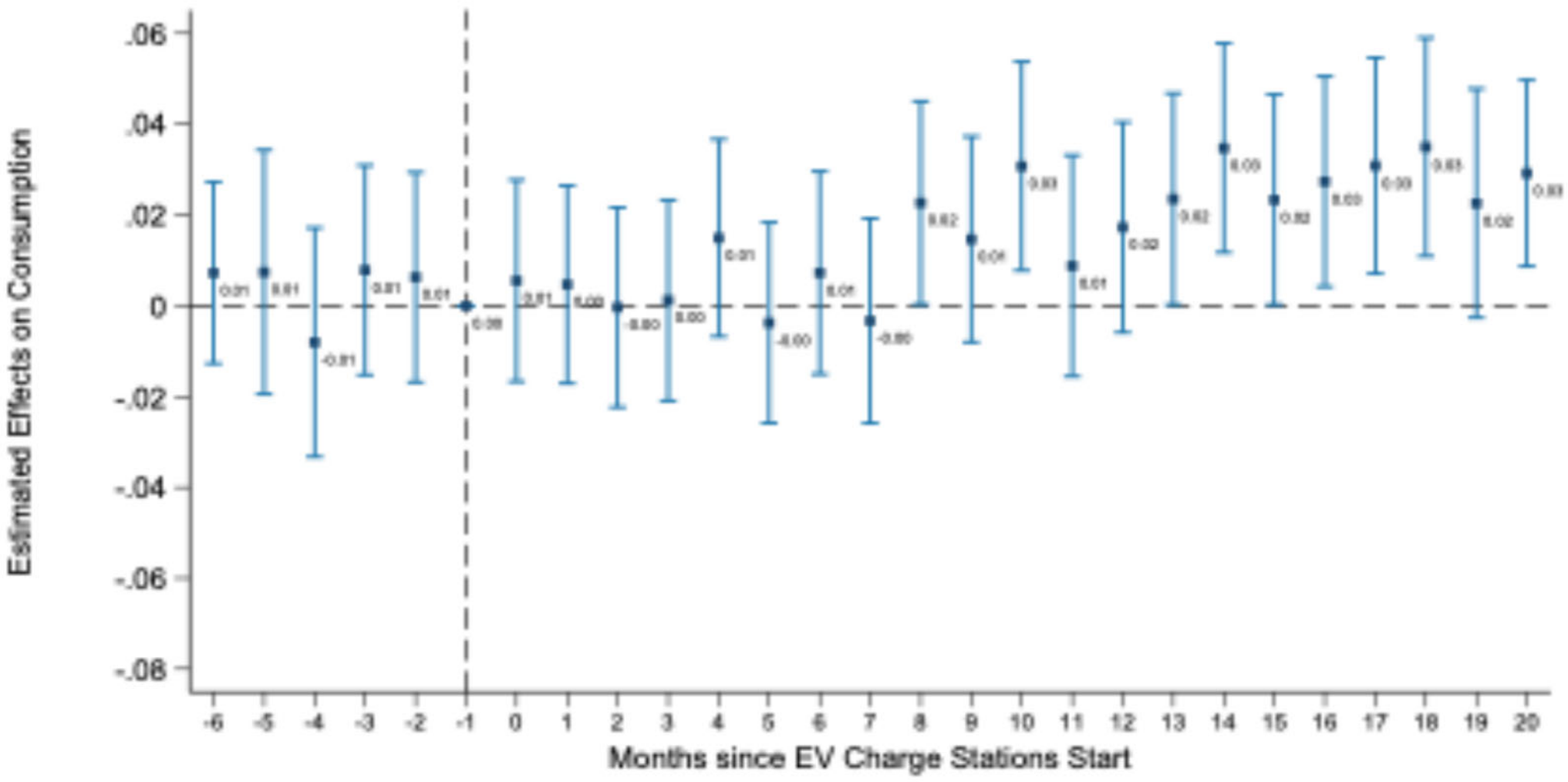

**Figure 1:** Event-Study on Household Consumption

Notes: The figures present an event-study examining the impact of nearby EV charging stations on total household consumption. Dependent variable is the logarithmic value of monthly consumption from January 2019 to March 2021. Consumption is constructed by aggregating across payment types after trimming the top and bottom 1% of expenditures within each type. We exclude inactive accounts with more than one month of zero consumption. Individual, time, and planning area-by-year fixed effects are included. Standard errors are clustered at the individual and month level. *$P < 0.10$; **$P < 0.05$; ***$P < 0.01$; Error bars indicate 95% confidence intervals.

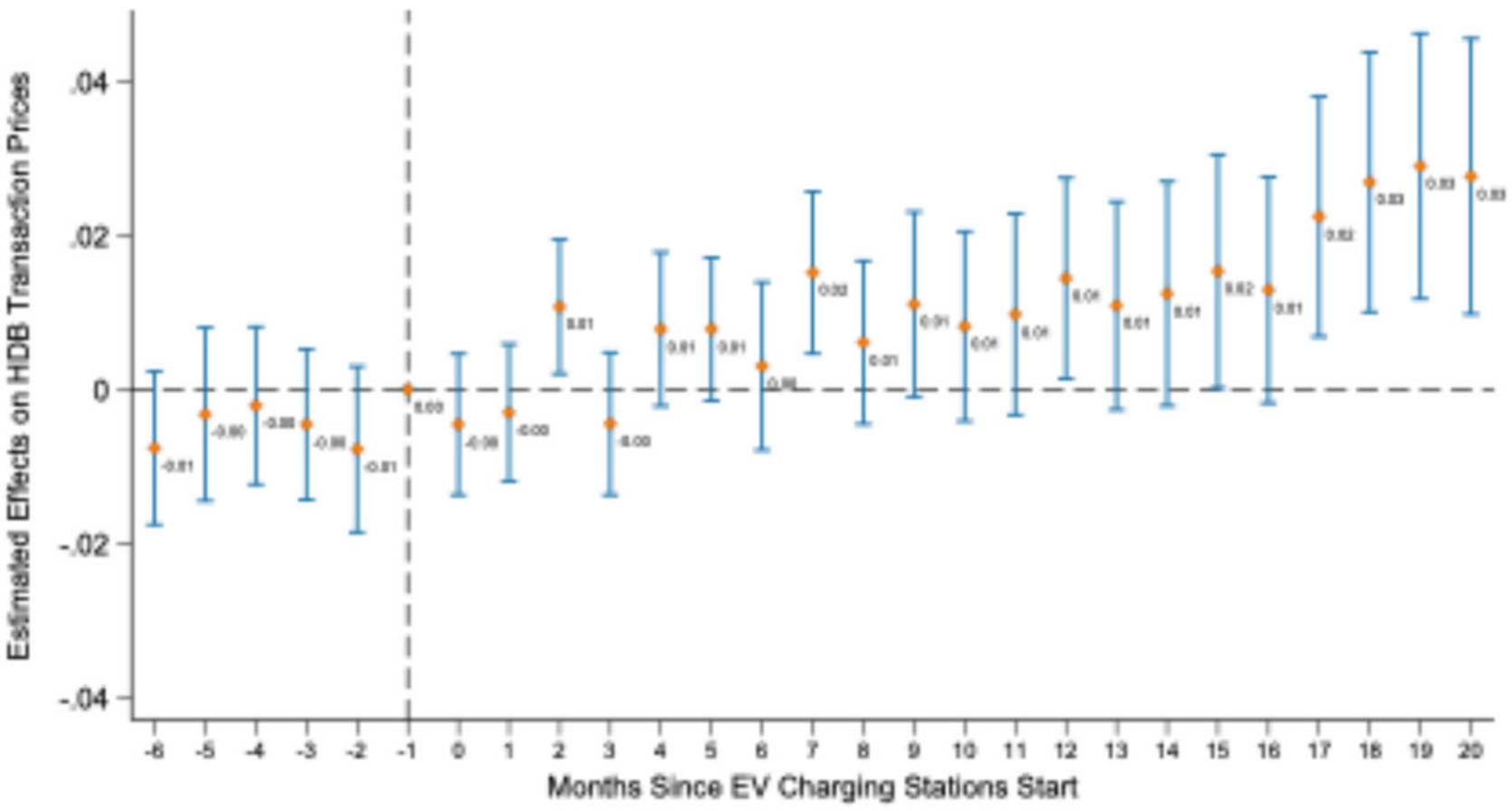

a. Public Housing

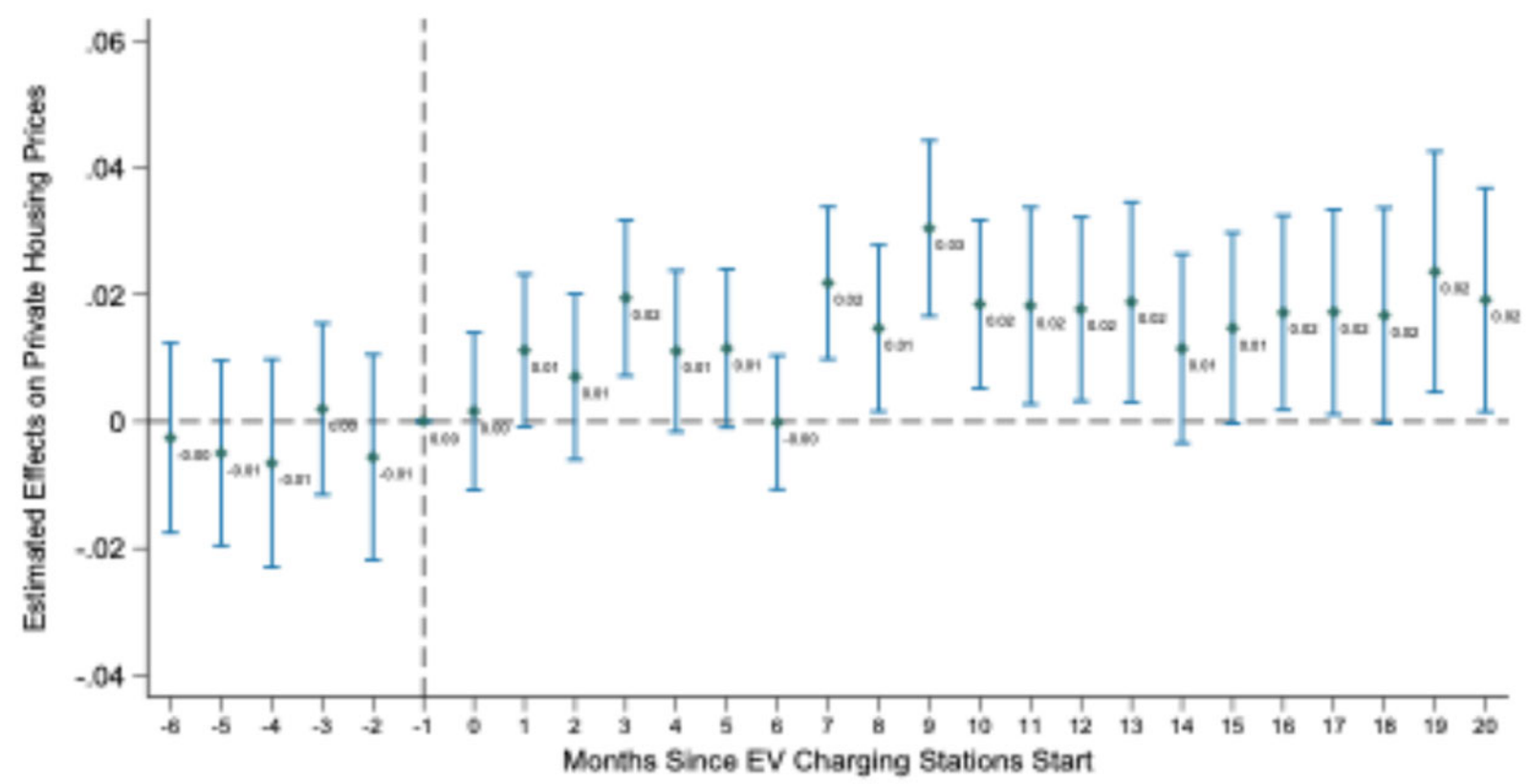

b. Private Housing

**Figure 2:** Event-Study on Housing Price

Notes: The figures present event-study estimates of the impact of nearby EV charging stations on housing prices for public and private housing, separately. Dependent variable is the logarithmic value of transaction price from January 2019 to March 2021. Charger-by-building, time, and region-by-year fixed effects are included. Standard errors are clustered at the postal code level. *$P < 0.10$; **$P < 0.05$; ***$P < 0.01$; Error bars indicate 95% confidence intervals.

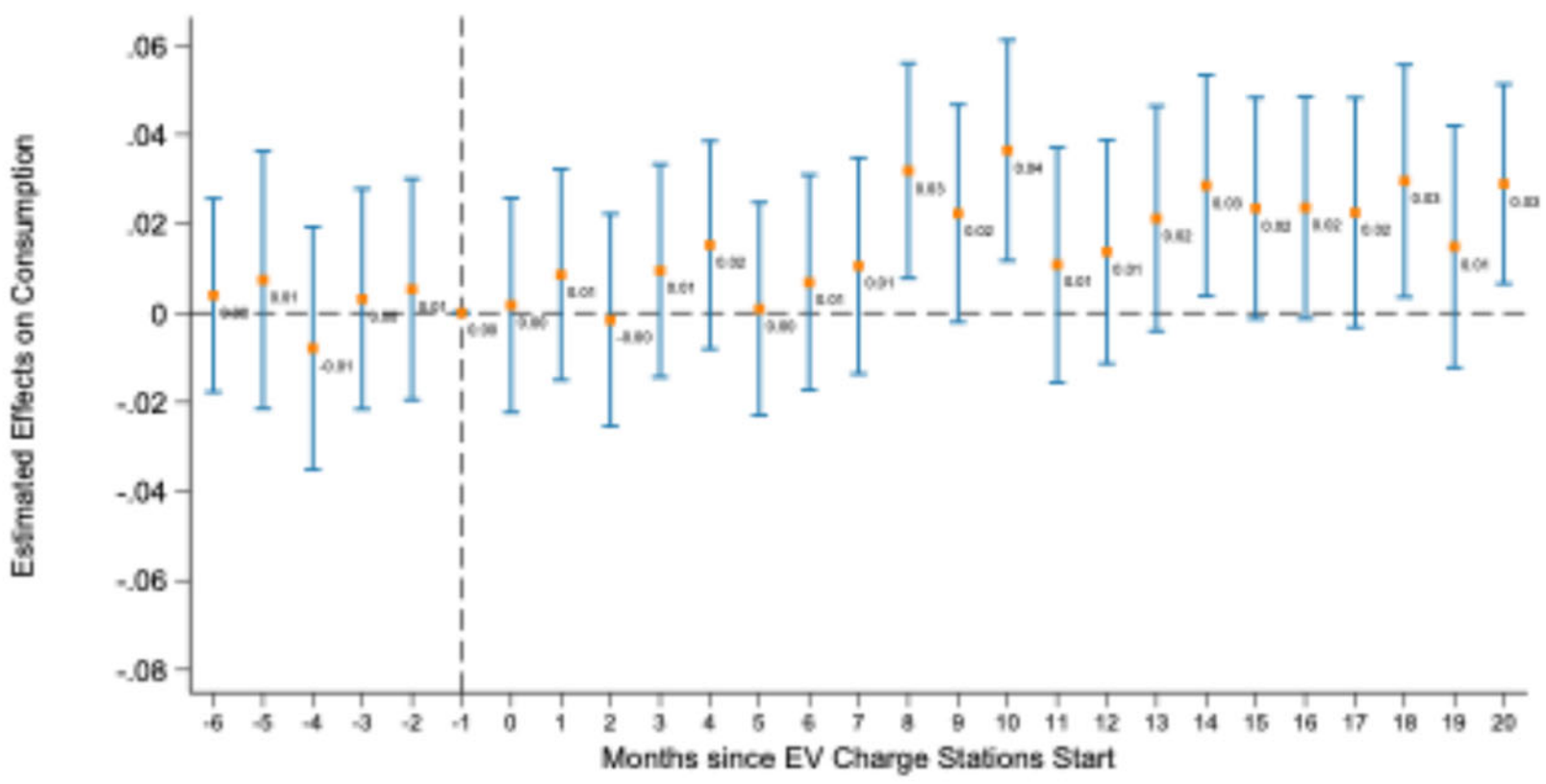

a. Public Housing

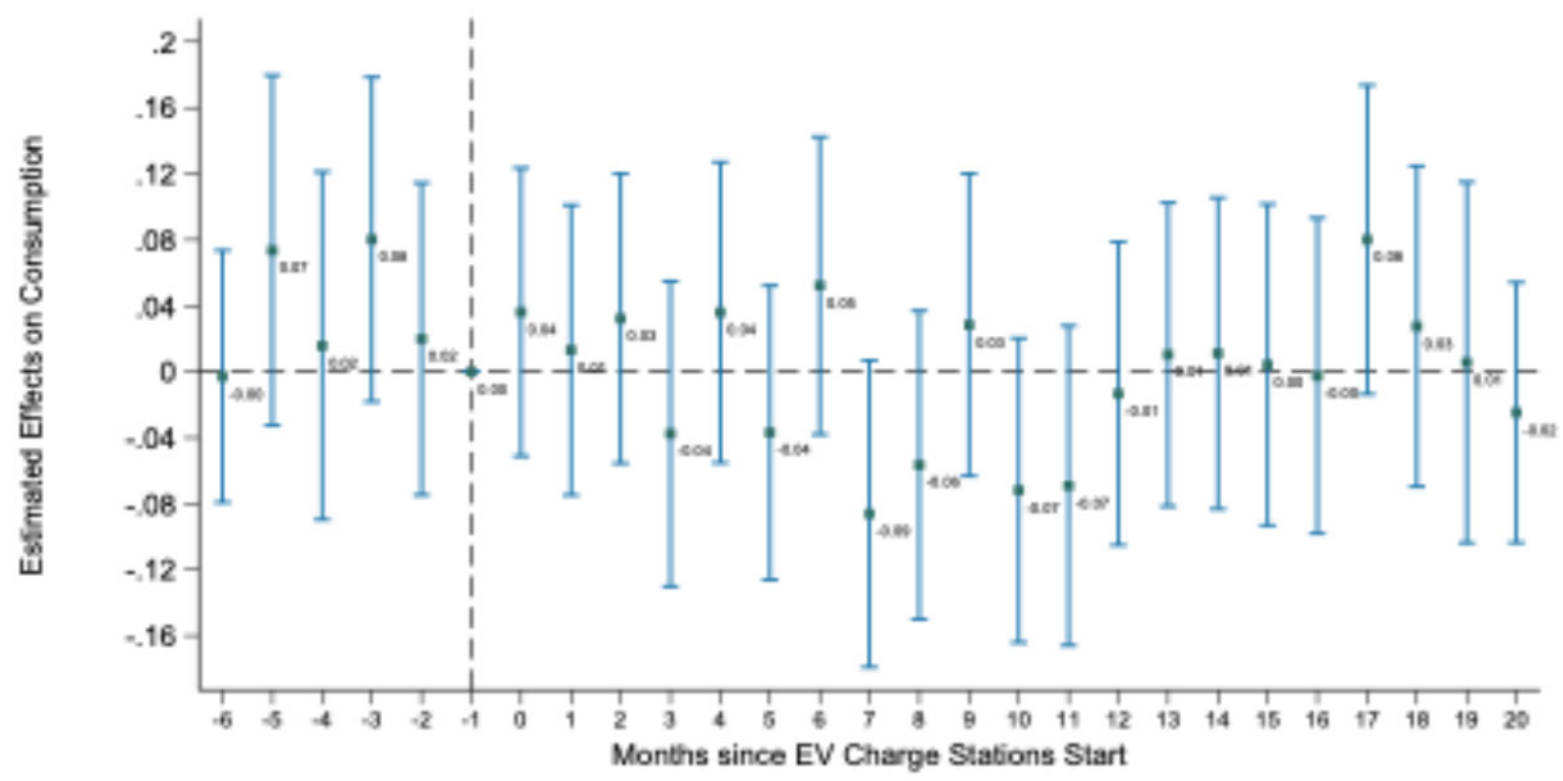

b. Private Housing

**Figure 3:** Event-Study on Household Consumption by Housing Type

Notes: The figures present an event-study examining the impact of nearby EV charging stations on total household consumption for public and private housing, separately. Dependent variable is the logarithmic value of monthly consumption from January 2019 to March 2021. Consumption is constructed by aggregating across payment types after trimming the top and bottom 1% of expenditures within each type. We exclude inactive accounts with more than one month of zero consumption. Individual, time, and region-by-year fixed effects are included. Standard errors are clustered at the individual and month level. *P < 0.10; **P < 0.05; ***P < 0.01; Error bars indicate 95% confidence intervals.

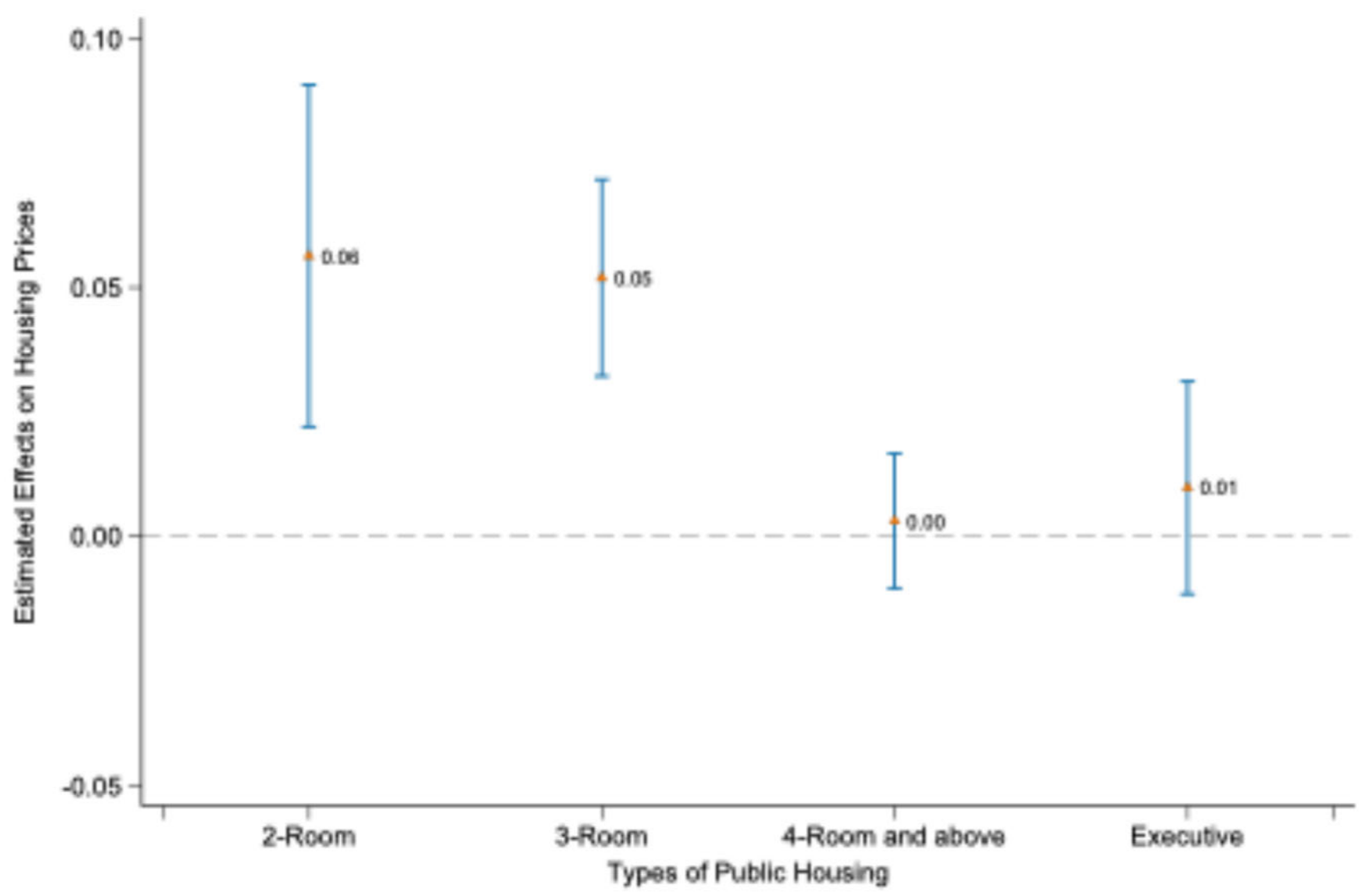

**Figure 4:** Heterogeneous Effects of Different Types of Housing

Notes: The figures show how the impact of nearby EV charging stations on public housing price varies across housing types, defined by the number of rooms. Dependent variable is the logarithmic value of transaction price from January 2019 to March 2021. Charger-by-building, time, and region-by-year fixed effects are included. Standard errors are clustered at the postal code level. *P < 0.10; **P < 0.05; ***P < 0.01; Error bars indicate 95% confidence intervals.

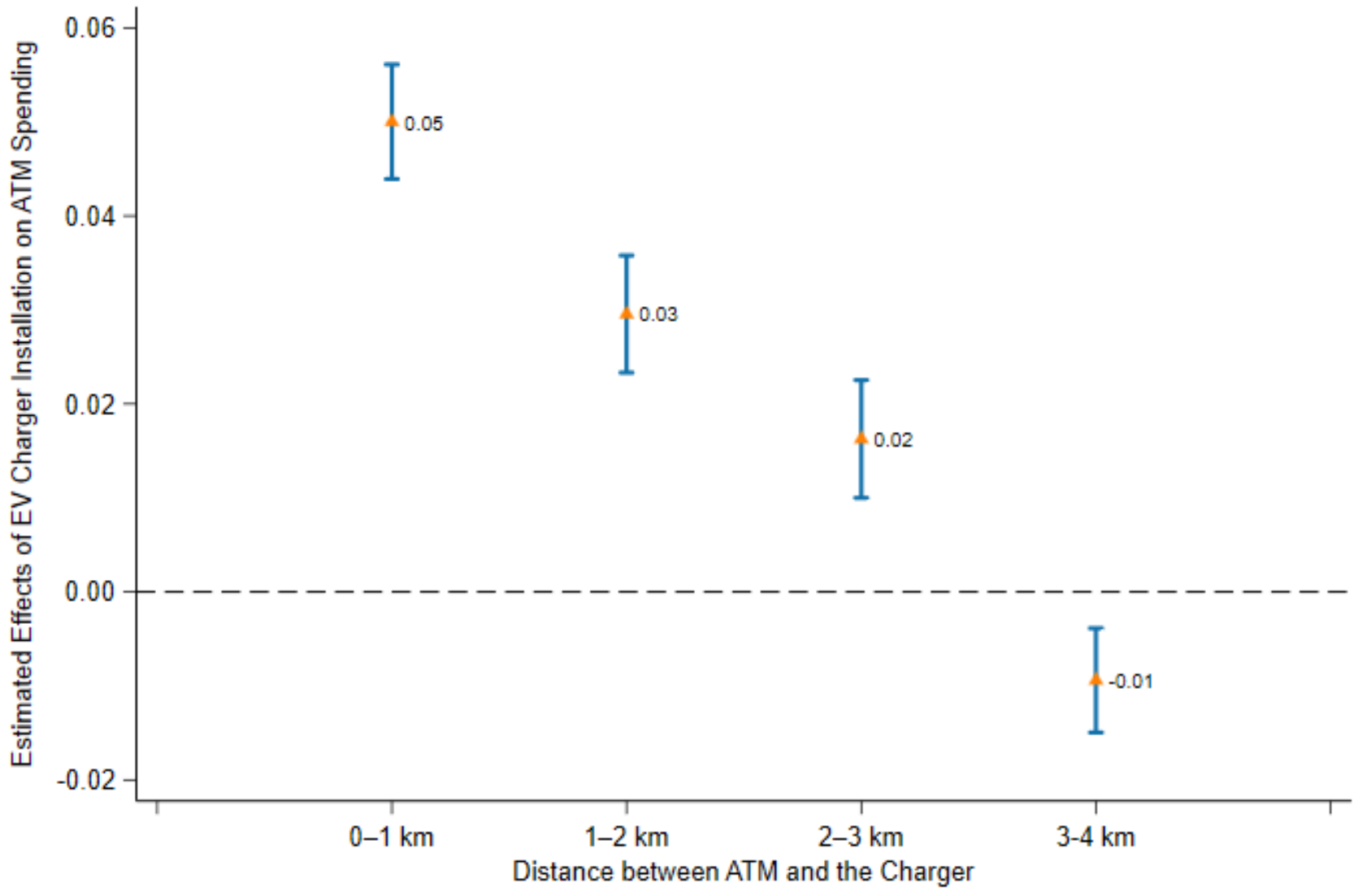

b. Public Housing

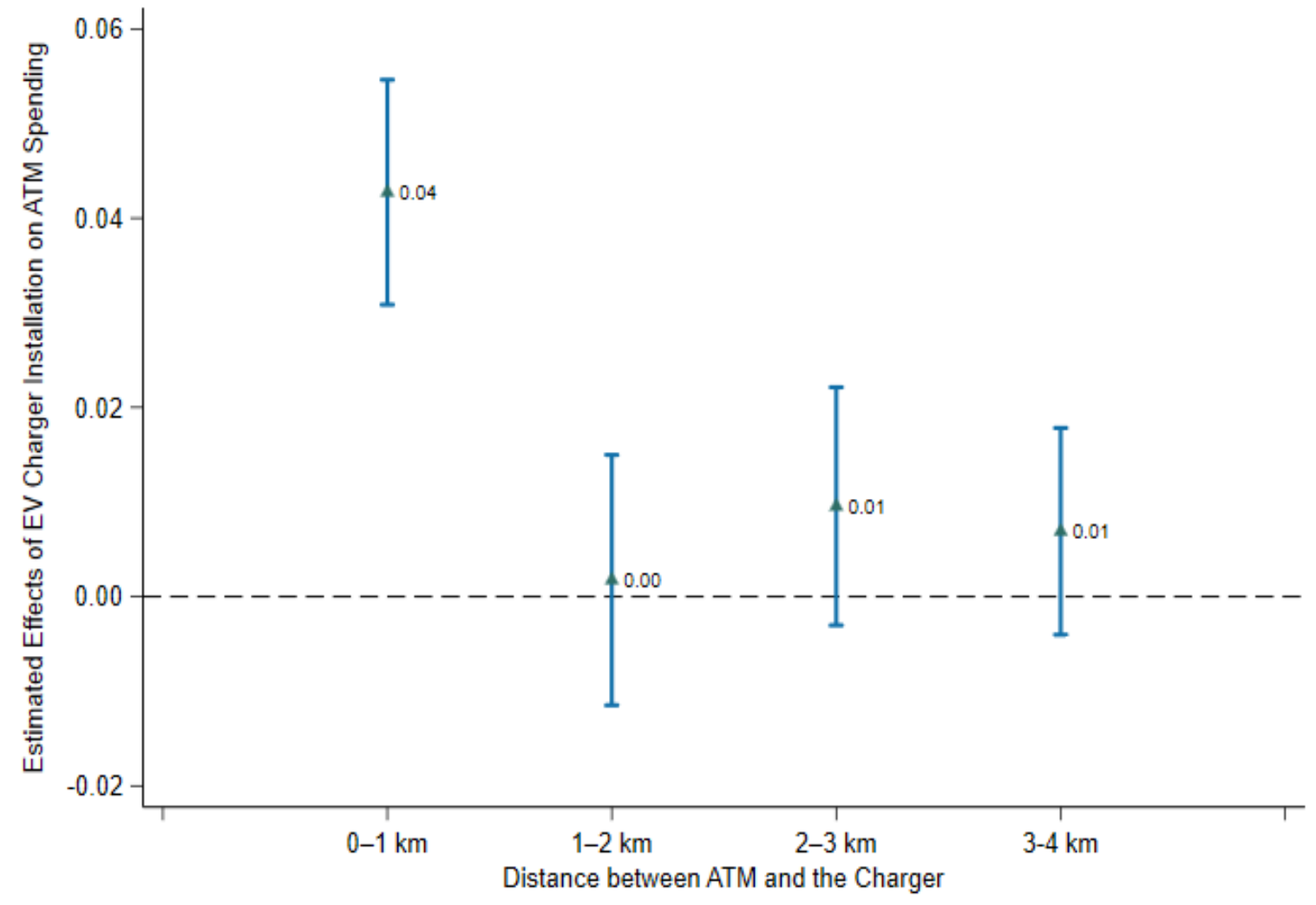

b. Private Housing

**Figure 5:** Heterogeneous Effects of Distance to the Nearest Charger on ATM Spending

Notes: This figure shows the estimated effects of EV charger installation on ATM cash withdrawals by the distance between the ATM and the nearest EV charger. Effects are reported relative to a control group of ATMs located at least 4 kilometers away from the EV charger. Dependent variable is the logarithmic value of monthly consumption by ATM after removing the top and bottom 1% outliers from January 2019 to March 2021. We exclude inactive accounts with more than one month of zero consumption. Individual, time, and planning area-by-year fixed effects are included. Standard errors are clustered at the individual and month level. *P < 0.10; **P < 0.05; ***P < 0.01; Error bars indicate 95% confidence intervals.

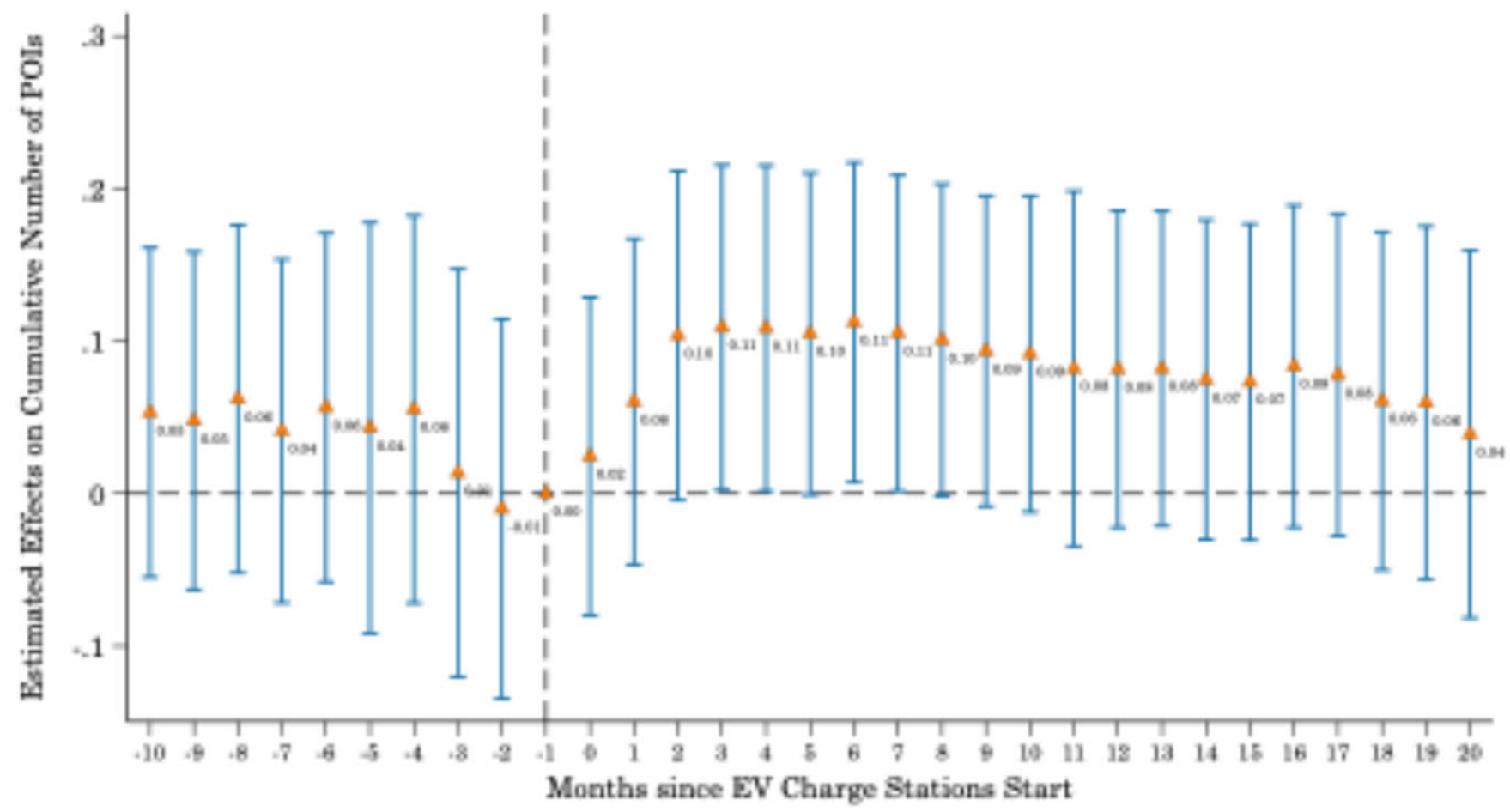

**Figure 6:** Event Study of EV Charger Installations and New POIs

Notes: This figure shows the event study of EV charger installations on newly built POIs in the treated postal code zones. Dependent variable is the monthly cumulative number of POIs in 4-digit postal code zones from January 2019 to March 2021. Postal code zone, time, and planning area-by-year fixed effects are included. Standard errors are clustered at the zone and month levels. *P < 0.10; **P < 0.05; ***P < 0.01; Error bars indicate 95% confidence intervals.